\documentclass[11pt,twoside]{article}
\usepackage{asp2010}

\newcommand{\reaction}[6]{\nuc{#1}{#2}(#3,#4)\/\nuc{#5}{#6}}
\newcommand{\nuc}[2]{\ensuremath{^{#1}}#2}

\resetcounters

\begin{document}
\label{authorguide}

\title{White dwarf mergers and the origin of R~Coronae Borealis stars}

\author{P.~Lor\'{e}n--Aguilar$^{1,2}$,
        R.~Longland$^{3,2}$, 
        J.~Jos\'{e}$^{3,2}$, 
        E. Garc\'{\i}a--Berro$^{1,2}$,
        L.~G.~Althaus$^{4}$, \& 
        J.~Isern$^{5,2}$}

\affil{$^1$Departament de F\'{\i}sica Aplicada, 
           Universitat Polit\`{e}cnica de Catalunya, 
           c/Esteve Terrades, 5,
           08860 Castelldefels, 
           Spain\\
       $^2$Institut d'Estudis Espacials de Catalunya, 
           c/Gran Capit\`{a} 2-4, 
           Ed. Nexus 201,
           08034 Barcelona, 
           Spain\\
       $^3$Departament de F\'{\i}sica i Enginyeria Nuclear,
           Universitat Polit\`{e}cnica de Catalunya, 
           c/Comte d'Urgell 187,
           08036 Barcelona, 
           Spain\\
       $^4$Facultad de Ciencias Astron\'omicas y Geof\'{\i}sicas,
           Universidad Nacional de La Plata, 
           Paseo del Bosque s/n, 
           (1900) La Plata, 
           Argentina\\
       $^5$Institut de Ci\`encies de l'Espai (CSIC), 
           Facultat de Ci\`encies,
           Campus UAB, 
           08193 Bellaterra, 
           Spain}

\begin{abstract}
We present a nucleosynthesis study of the merger of a $0.4\, M_{\sun}$
helium white dwarf with  a $0.8\, M_{\sun}$ carbon-oxygen white dwarf,
coupling the thermodynamic  history of Smoothed Particle Hydrodynamics
particles  with  a   post-processing  code.   The  resulting  chemical
abundance  pattern,  particularly  for  oxygen  and  fluorine,  is  in
qualitative  agreement  with  the  observed abundances  in  R  Coronae
Borealis stars.
\end{abstract}

\section{Introduction}

Hydrogen-deficient  stars  with   high  carbon  abundances  (with  the
exceptions of  PG~1159 and  CSPN WR stars)  can be divided  into three
categories  according   to  their  effective   temperatures:  Hydrogen
Deficient  Carbon (HdC) stars,  R~Coronae~Borealis (RCB),  and Extreme
Helium  (EHe) stars.   They are  often considered  to be  at different
evolutionary stages following a  common origin \citep{PAN04}, which is
of particular interest, owing to their unique surface composition.  As
well  as  being  enriched  in  carbon and  oxygen  relative  to  solar
abundances,  they  also exhibit  enrichment  in  N,  Ne, F,  and  some
s-processed  material.  This  abundance pattern  is hard  to reconcile
with their initial  composition, requiring alternative scenarios.  Two
theories  are the  Final  Flash (FF)  and  the Double-Degenerate  (DD)
scenarios.  In  the DD  scenario, the favored  one, the merger  of two
white dwarfs (He+CO), with a  total mass below the Chandrasekhar limit
gives rise to a single object whose composition is an admixture of the
two  white dwarfs  \citep{ASP00}. Within  this scenario  the following
question  can be  raised:  does nuclear  processing  occur during  the
merging event?  If  no nuclear burning occurs ---  the ``cold'' merger
hypothesis --- the observed  abundances might be reproduced by partial
mixing of material previously processed during the AGB phase of the CO
white dwarf  progenitor.  However,  recent calculations based  on this
scenario failed to match the observed abundance pattern \citep{JEF11}.
Another  possibility  involves  a   ``hot''  merger  event,  in  which
significant nuclear processing occurs.  Indeed, \cite{CLA07} suggested
that this  scenario may  account for the  high abundances  of $^{18}$O
observed  in some  stars.  Here  we present  a detailed  study  of the
nucleosynthesis accompanying white dwarf mergers.

\section{Initial models}
\label{sec:SPH}

The models used as input for the nucleosynthesis calculations reported
here are those of \cite{GUE04} and \cite{LOR09}.  In these simulations
the final remnant of the merger  consists of a central hot white dwarf
containing almost the  entire primary.  On top of it  a hot corona can
be found, which  contains most of the mass  of the disrupted secondary
and  a  small admixture  of  the  primary.   Finally, surrounding  the
compact remnant, a rapidly rotating  disk is formed.  The ejected mass
is  very small. The  nucleosynthesis reported  here is  calculated for
tracer particles  using a 327 nucleus  post-processing network ranging
from  H  to Ga.   The  reaction rates  are  adopted  from the  REACLIB
database  \citep{REACLIB}  with  updates  on experimental  rates  from
\cite{ILI10a}.  SPH  particles that  reside in the  hot corona  of the
final object  characterize the  ``atmosphere'' observed in  a hydrogen
deficient star.  The tracer particles used to follow the thermodynamic
history of the merger are  therefore picked from $0.005 < R/R_{\sun} <
0.05$.  This  range includes particles that  represent everything from
the  surface of  the central  dense object  to the  inner edge  of the
accretion disk.   A total of  10\,000 tracer particles have  been used
here to ensure a representative sample.

\begin{figure} 
\centering
  \includegraphics[width=0.7\hsize]{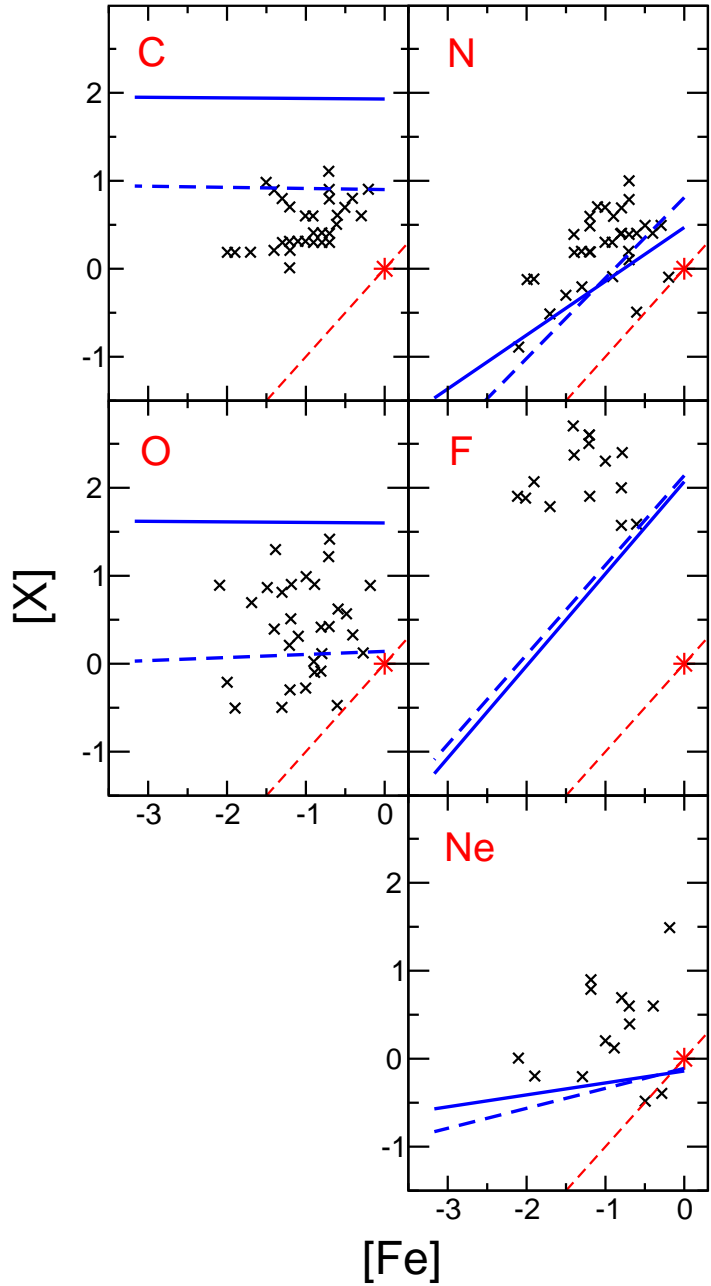}
  \caption{Surface  abundance determinations  taken  from \cite{JEF11}
           compared to our yields.  We show the logarithmic abundances
           relative  to solar,  [X], for  individual elements  and for
           iron,  [Fe].  Solid  lines  are our  calculations  assuming
           ``deep''  convective  mixing, while  dashed  lines are  for
           ``shallow''  mixing.   The  asterisk corresponds  to  solar
           abundances, while  the  lines intersecting solar abundances
           are the values scaled with metallicity.}
  \label{fig:results}
\end{figure}

Our simulations consider the existence  of thin He and H shells, which
are expected  to survive prior evolution.   Therefore, three different
regions  (CO-, He-, and  H-rich) are  defined in  the CO  white dwarf,
while two (He- and H-rich) are  adopted for the He white dwarf.  These
regions are  defined according to  the mass contained.  We  follow the
notation of \cite{SAI02} --- who referred  to the mass of the He shell
in a CO white dwarf as $M_{\textrm{He:CO}}$. Using this convention the
different shells are $M_{\textrm{CO:CO}} = 0.78$, $M_{\textrm{He:CO}}=
0.019$,  $M_{\textrm{H:CO}}  =  0.001$,  $M_{\textrm{He:He}}=  0.399$,
$M_{\textrm{H:He}} = 0.001$ \citep{REN10}. In our calculations we have
employed solar  metalicity and $Z=10^{-5}$. Finally,  although the SPH
calculations  suggest that the  entire hot  corona will  be convective
\citep{Rea11}, we  adjust the depth of  mixing to study  its impact on
the final  abundances. Specifically,  we consider two  cases, ``deep''
mixing, in which  everything in the range $0.005  < R/R_{\sun} < 0.05$
is  mixed homogeneously,  and ``shallow''  mixing, in  which  we adopt
$0.014 < R/R_{\sun} < 0.05$.

\section{Results and Discussion}
\label{sec:nucleosynthesis}

The  mass  averaged  abundances  obtained  from  the  tracer  particle
nucleosynthesis are  compared with  the observed abundances  in figure
\ref{fig:results}.   Carbon  depends  strongly  on  the  mixing  depth
assumed in  the hot  corona of  the merger product  but appears  to be
consistently high in our calculations. Nevertheless, the overall trend
of  carbon  abundance  with  respect  to metalicity  agrees  with  the
observational data.  The  $^{13}$C$/^{12}$C ratio for solar metalicity
is $\sim  2 \times 10^{-5}$,  consistent with the predicted  nature of
He-burned material,  but considerably higher than  the observed ratio.
This  disagreement could  arise  from the  limited  number of  nuclear
species adopted in our calculations.  The oxygen abundance also agrees
fairly well  with the observed data,  showing a large  range of values
depending on  the mixing depth assumed. This  dependence could explain
the  observed  scatter in  oxygen  in EHe  and  RCB  stars, where  the
observed oxygen abundance  could depend strongly on the  nature of the
individual merging  event.  This  picture is not,  however, consistent
with the relatively low scatter  in carbon abundances.  With regard to
the  isotopic abundances,  our  solar metalicity  model  for deep  and
shallow   mixing   yields    $^{16}$O/$^{18}$O~$=   370$   and   $19$,
respectively.   The $^{18}$O has  clearly been  enhanced in  the outer
regions  of the  hot corona  with respect  to solar  abundances (where
$^{16}$O/$^{18}$O~$\approx 400$), although not  to the extent found in
some HdC and RCB stars \citep{CLA07}.

Nitrogen  follows the expected  trend indicating  that it  is enhanced
only through CNO cycling in the parent stars.  Although it is slightly
reduced  by the  \reaction{14}{N}{$\alpha$}{$\gamma$}{18}{F} reaction,
it  is  present  in  such  a  high concentration  that  it  is  barely
diminished in  the short time-scales involved.   The overabundances of
fluorine obtained in our models are high, but not high enough to fully
agree with the  observational data. The lack of  sensitivity to mixing
depth is an indication that  fluorine is produced homogeneously in the
merging  event,  and not  in  any  particular  region of  the  merger.
Enrichment of  fluorine in our  models is a great  success, suggesting
that hydrogen deficient stars could be the result of a hot white dwarf
merger  event  since  it   is  challenging  to  reconcile  those  high
abundances with a  cold merger event.  Finally, the  abundance of neon
following the  merger is  consistent with the  metallicity, indicating
that it  is not processed  in either the  merger event, or  during the
preceding evolution of the parent stars.

\section{Conclusions}

The results presented in this work show that nuclear processing during
the merger  event can account for  some of the  observed abundances in
these  stars,  particularly  for  fluorine,  which  cannot  be  easily
synthesized in  the cold merger  scenario.  There are,  however, still
open  questions  regarding   the  expected  nucleosynthesis:  neon  is
difficult  to  produce at  the  observed  levels,  while the  absolute
abundance of carbon is too high in our models.

\acknowledgements

This  work  has  been   partially  supported  by  the  Spanish  grants
AYA2010-15685,  AYA08-1839/ESP,  and  AYA2008-  04211-C02-01,  by  the
E.U. FEDER funds, and by the ESF EUROCORES Program EuroGENESIS through
the MICINN {grants EUI2009-04167 and 04170}.

\bibliography{PaRi}

\end{document}